# Revisiting π Backbonding: The Influence of *d* Orbitals on Metal-CO Bonds and Ligand Red Shifts

Daniel Koch,[a] Yingqian Chen,[a] Pavlo Golub[a] and Sergei Manzhos[*,b]

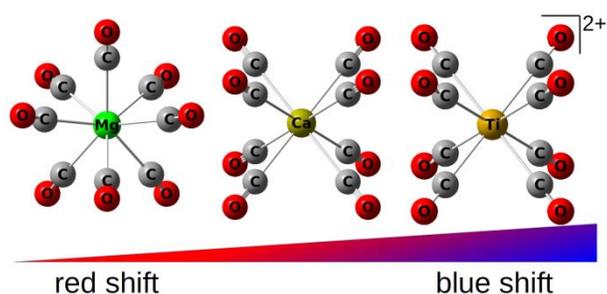

**Synopsis:** The influence of metal *d* functions on complex stability, C-O bond strength, real-space charge transfer, and bond order in carbonyl complexes of Mg, Ca and Ti is investigated from first principles. Stronger C-O bonds are found if metal *d* functions are present, contrary to what would be expected based on the commonly employed π backbonding picture.

[a] Department of Mechanical Engineering, National University of Singapore, 9 Engineering Drive 1, Singapore 117575

[b] Centre Énergie Matériaux Télécommunications, Institut National de la Recherche Scientifique, 1650, Boulevard Lionel-Boulet, Varennes QC J3X1S2 Canada, Tel.: +1 514 2286841, Email: sergei.manzhos@emt.inrs.ca

**Abstract:** The concept of π backbonding is widely used to explain the complex stabilities and CO stretch frequency red shifts of transition metal carbonyls. We theoretically investigate a non-transition metal 18-electron carbonyl complex (Mg(CO)$_8$) and find a pronounced CO red shift without metal-carbon π bonds. Moreover, we use truncated basis sets on the "honorary" and true transition metals Ca and Ti in Ca(CO)$_8$ and [Ti(CO)$_8$]$^{2+}$ complexes to probe the influence of $d$ functions on carbonyl complex stability, C-O bond strength, metal-to-ligand charge transfer and bond order compared to hypothetical complexes without metal-$d$ contributions. We find that the occurrence of metal-ligand π bonds through metal $d$ functions greatly enhances the complex stabilities on one hand but only slightly affects the CO red shift on the other hand. This does not correspond to the classical rationalization of transition metal-CO bonds as synergistic σ donation/π backdonation.

## 1. Introduction

The notion of donor and acceptor bonds is a ubiquitous concept appearing in many areas of chemistry and is widely used for the rationalization of otherwise not easily understandable chemical properties, especially of complex compounds.[1, 2] Using an *ad hoc* concept of frontier orbital interactions, it lays out the foundation for more specific models like the one proposed by Dewar, Chatt and Duncanson[3, 4] for the binding between alkene ligands and transition metal (TM) centers or π backbonding between TM atoms and carbonyl (CO) ligands. To this day, the common consensus for the TM-CO bonding picture remains the synergistic donation of CO electrons via σ bonds into empty TM $d$ orbitals and backdonation from the metal center into π* orbitals of the CO ligands, mitigated by M-C π bonds due to the large overlap of TM $d$ and CO π* orbitals. The backbonding into the unoccupied CO π* orbitals is also commonly used to explain the CO bond destabilization and hence CO stretch frequency red shift of so-called "classical" carbonyl complexes, representing the majority of TM carbonyl complexes (although for non-classical complexes the π backbonding is just said to be of less importance than the primary σ bond, resulting in no red-shift or even leading to a blue shift[5]). This traditional picture is very well investigated and supported by a large number of quantum chemical studies published over the course of decades, employing different computational methods on different conceptual levels and for a wide variety of TM carbonyls.[6-12] However, analyses of the bond character in TM carbonyls are often based on concepts and measures operating with the use of molecular orbitals (MOs), non-unique sets of one-electron wave functions which do not directly correspond to any physical observable (and furthermore often built up from atomic one-electron basis functions, which is not necessarily a good basis for the compound), a point previously raised by Cortes-Guzman and Bader[12] who rather based their discussion of the TM-CO bond on the topological analysis of electronic density in real space.

Another common concept to classify bonding in complex compounds based on the analysis of molecular orbitals is the one of formal oxidation states (FOS). CO is commonly classified as L-type ligand, meaning that the TM-CO bond is treated as completely dative in nature and no electron transfer in the formal oxidation state framework is taking place.[13] However, this concept does not necessarily reflect real-space electron density distribution, leading to situations in which a significant depletion of electron density around the metal center can take place (and in turn an increase around the C and O nuclei), although formally (in the FOS picture) no charge is transferred. Since the internal energy of the complex in turn depends on the specific electron density distribution, it is this real-space charge transfer to which the differences in chemical and physical properties (complex stabilities, CO stretch frequencies etc.) of carbonyl compounds can be assigned. The correlation between CO stretch frequencies and transferred real-space charge to the ligands as well as the stabilizing effect of this transfer have been previously reported in literature.[12, 14]

Interestingly, in a recent investigation, Wu *et al*. were able to isolate the octacarbonyl complexes of Ca, Sr and Ba in a low-temperature neon matrix.[7] Although not $d$ block elements themselves, these "honorary" transition metals were found to form compounds whose many-body wave functions have large admixtures of configuration state functions with $d$ occupations due to a strong static correlation.[15] This influence of $d$ contributions to the wave function of higher group II elements provides an elegant way to understand the formation of their carbonyl complexes in the existing framework of MO theory and π backbonding, normally confined to $d$ block element complexes. In this work, we investigate the influence of $d$ contributions to electronic structure, binding energies and CO red shift of a honorary TM carbonyl exemplified by $Ca(CO)_8$ and compare it with theoretically stable non-TM ($Mg(CO)_8$) and TM ($[Ti(CO)_8]^{2+}$) 18-electron carbonyl complexes under the same approximations. We chose the Mg carbonyl complex since the $d$ states of the Mg central atom are far above the atomic HOMO level and hybridization with Mg $d$ states is expected to be insignificant for the occupied orbitals of the carbonyl complex as well, whereas for Ca these states lie significantly lower and large $d$ contributions to the Ca(CO)$_8$ frontier orbitals are expected (and have been reported[7]). For Ti$^{2+}$ on the other hand, the ground state valence configuration is purely $3d^2$ and $d$ orbital contributions to the frontier orbitals of [Ti(CO)$_8$]$^{2+}$ are expected to be very pronounced, providing a reference for the high $d$ admixture case. We use a straightforward orbital-independent basis set truncation approach to theoretically probe the role of $d$ functions on the metal-CO bonds in these three systems with regard to complex stability, C-O bond strength, real-space charge transfer, and bond order. For the last two properties, we utilize Bader analysis, a common approach to determine the charge of atoms in molecules by integration of electron density basins associated with them, and delocalization indices (DIs), a measure for the delocalization of electrons between pairs of these basins and indicative of the covalent character of the bonds connecting them. For details on these two approaches, see References 16-20. We find that the omission or lack of $d$ contributions destabilizes the carbonyl complexes, while making M-CO bonding more ionic and the CO stretch frequency red shift more severe, in contrast with the common π backbonding-based explanation of TM-CO bonding.

## 2. Results and Discussion

Electronic structure computations, geometry optimizations and vibrational analyses of the isoelectronic complexes Mg(CO)$_8$, Ca(CO)$_8$ and [Ti(CO)$_8$]$^{2+}$ were performed with the program package *Gaussian 16*[21] using the M06-2X hybrid density functional[22] in spin-unrestricted calculations, in vacuum. A correlation-consistent Dunning *cc*-pVQZ basis set[23-26] was used on all atoms, since this basis set type includes higher angular momentum basis functions as polarization components which allows to probe the eventual influence of $d$ functions even for elements without valence $d$ orbitals. Total and binding energies were found to be reasonably converged at this basis set size, for details see the section *Basis Set Test* in the Supporting Information. Convergence thresholds were set to $10^{-6}$ $E_h$ for the total energy and $1.5 \cdot 10^{-5}$ $E_h/a_0$ for the interatomic forces. In accordance with Reference 7 the structures were initialized in cubic ($O_h$) and square antiprismatic ($D_{4d}$) geometry in triplet and singlet spin state, respectively, to find the correct ground state geometry. For comparison between different complexes, the average CO stretch frequencies were calculated from the eight CO stretching modes, whose reduced masses were found to deviate not more than 1% from that of free CO and were hence considered as sufficiently decoupled from the other degrees of freedom. Values for the CO stretch frequencies will be given as difference to the calculated CO stretch frequency of free carbon monoxide, $v_0$, which was found at 2279 cm$^{-1}$ (exp.[27]: 2143 cm$^{-1}$) and with a C-O bond length of 1.1193 Å (exp.[28]: 1.1282). The sometimes used scaling of calculated carbonyl frequencies by the ratio of free CO stretch frequency values from experiment and theory[29] was omitted here, since we are only interested in the relative qualitative behavior of the CO stretch frequencies and not in comparisons to experimental values. To determine the influence of the metal *d* orbitals, the basis set on the central atom was successively truncated by removal of orbital sets with different radial extent and nodal structure from the basis set definition of Ca, Mg and Ti for the calculation of ground state energies, frequencies and electronic structure. The Bader charges and delocalization indices were computed with the *Dgrid*-4.7 program.[30]

We restrict ourselves to the detailed discussion of 18-electron octacarbonyl complexes, since these were found to be electronic potential energy surface minima with Mg, Ca and Ti$^{2+}$ (even under basis set alterations) with our computational setup and also present more realistic model systems for this investigation than e.g. a simplistic single MCO unit which would not account for the effects of several ligands.

We also tested for the effect of dispersion corrections (via DFT-D3 correction[31]) and basis set superposition errors[32, 33] (BSSE) and found both to be inconsequential for the conclusions drawn in this work. For details see section *Dispersive and Basis Set Superposition Effects* in the Supporting Information.

## 2.1 Molecular and Electronic Structure

The "honorary" and true TM complexes Ca(CO)$_8$ and [Ti(CO)$_8$]$^{2+}$ were found to exhibit a $^3O_h$ ground state (in accordance to Reference 7), while for Mg(CO)$_8$ the $^1D_{4d}$ structure was found to be slightly more stable with the full, and significantly more stable with truncated *cc*-pVQZ basis sets (see section *Influence of Complex Geometry* in the Supporting Information). The complex geometries are shown as insets in Fig. 2 - Fig. 4. It shall be noted at this point that the choice of stable octacarbonyl complex geometry ($^3O_h$ vs. $^1D_{4d}$) does not drastically influence the results. We have found the same trends across the less stable $O_h$ Mg(CO)$_8$, $D_{4d}$ Ca(CO)$_8$ and $D_{4d}$ [Ti(CO)$_8$]$^{2+}$ systems as described in the following sections for the $D_{4d}$ Mg(CO)$_8$, $O_h$ Ca(CO)$_8$ and $O_h$ [Ti(CO)$_8$]$^{2+}$ complexes. For details see section *Influence of Complex Geometry* in the Supporting Information. We employ the molar formation energy (in kJ/mol) $E_\text{f}$ of the octacarbonyl complexes as stability measure here, calculated as internal energy difference between the carbonyl complex and the metal atom/ion in its ground state configuration and eight CO molecules at infinite separation:

$$E_\text{f}([\text{M(CO)}_8]^{n+}) = E([\text{M(CO)}_8]^{n+}) - E(\text{M}^{n+}) - 8E(\text{CO}) \qquad (1)$$

The formation energies for the three investigated carbonyl complexes, as well as the average M-C and C-O bond lengths obtained after geometry optimization are listed in Table 1. For more detailed geometry specifications, see section *Complex Geometries* in the Supporting Information.

**Table 1** Complex geometries, metal Bader charges, CO stretch frequency changes (relative to free CO), complex formation energies and M-C/C-O delocalization indices for Mg(CO)$_8$ ($D_{4d}$), Ca(CO)$_8$ ($O_\text{h}$) and [Ti(CO)$_8$]$^{2+}$ ($O_\text{h}$) complexes obtained with M06-2X/*cc*-pVQZ and upon truncation of the metal basis sets after the *p* and *d* level.

|  | Mg(CO)$_8$ | | | Ca(CO)$_8$ | | | [Ti(CO)$_8$]$^{2+}$ | | |
| --- | --- | --- | --- | --- | --- | --- | --- | --- | --- |
|  | *sp* | *spd* | *spdfgh* | *sp* | *spd* | *spdfgh* | *sp* | *spd* | *spdfgh* |
| **r(M-C) [Å]** | 2.4170 | 2.3747 | 2.3715 | 2.6137 | 2.6003 | 2.6004 | 2.2941 | 2.4975 | 2.4969 |
| **r(C-O) [Å]** | 1.1271 | 1.1278 | 1.1279 | 1.1274 | 1.1262 | 1.1262 | 1.1159 | 1.1093 | 1.1093 |
| **q(M) [\|e\|]** | +1.65 | +1.68 | +1.68 | +1.75 | +1.45 | +1.45 | +3.49 | +1.69 | +1.69 |
| **$v_0$(CO)-v(CO)[cm$^{-1}$]** | -126 | -131 | -132 | -137 | -111 | -112 | -23 | +87 | +86 |
| **$E_\text{f}$ [kJ/mol]** | -10.8 | -26.4 | -28.5 | -145.0 | -310.1 | -313.5 | -577.2 | -1005.7 | -941.7 |
| **DI(M-C)** | 0.0420 | 0.0415 | 0.0412 | 0.0522 | 0.0787 | 0.0791 | 0.0777 | 0.1148 | 0.1105 |
| **DI(C-O)** | 0.8225 | 0.8182 | 0.8176 | 0.8305 | 0.8274 | 0.8272 | 0.8557 | 0.8755 | 0.8754 |

In the following, the MOs will be denoted by their irreducible representation in the $O_\text{h}$ or $D_{4d}$ geometry. Fig. 1 shows the MO schemes for the two investigated complex geometries, including the valence orbitals with the 18 electrons relevant for the M-CO interaction.

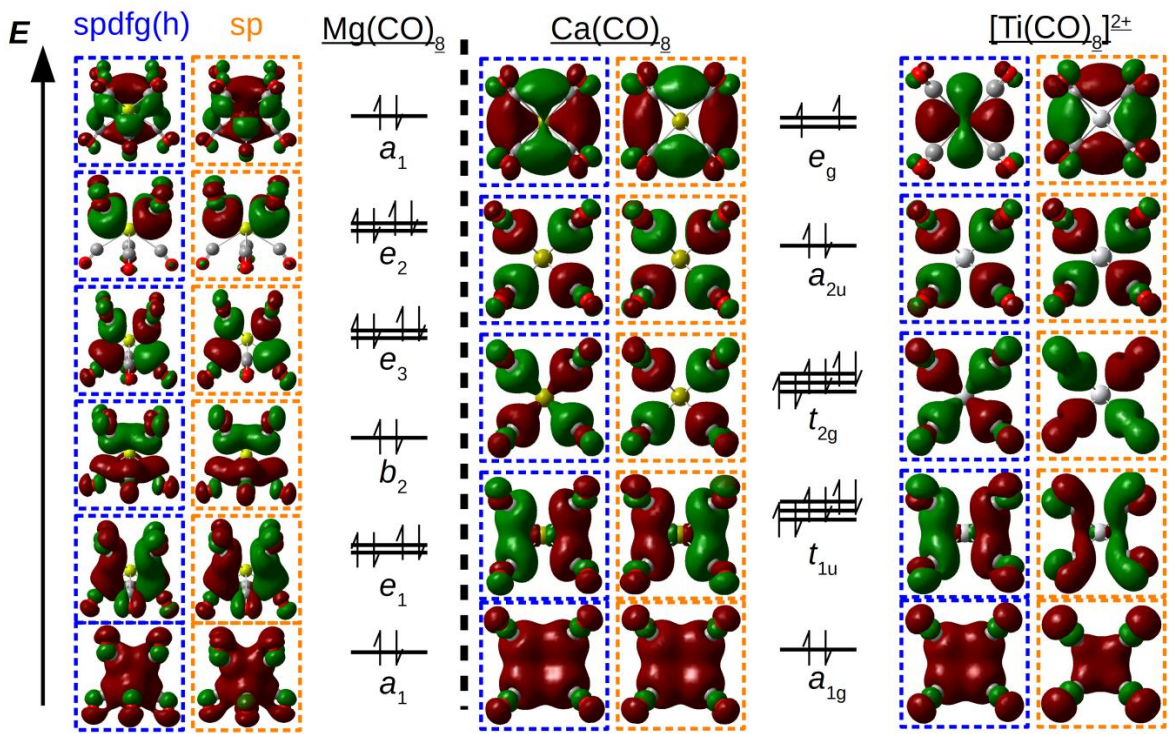

**Fig. 1** Outer valence MO schemes for the $D_{4d}$ complex $Mg(CO)_8$ (left) and the $O_h$ complexes $Ca(CO)_8$ and $[Ti(CO)_8]^{2+}$ (right). The insets show the corresponding isosurfaces (isovalue 0.03 $e^{1/2}/a_0^{3/2}$) of the orbitals, where the blue boxes (left columns) correspond to MOs with the full cc-pVQZ basis set with $s$, $p$, $d$ and higher polarization functions and the orange boxes (right columns) to the ones with a truncated basis containing only the full sets of $s$ and $p$ functions. In the cases of orbital degeneracies, only one of the orbitals is shown as isosurface plot to illustrate the main features of the corresponding MO.

Comparing the changes of MOs with decreasing $d$ character can help translate our findings into the framework of an orbital-based bonding picture. To avoid ambiguities associated with population and bond analysis techniques, we simply use the sum of squared coefficients $c_i^2$ of a certain basis function type $i$ in the investigated MO, divided by the total sum of all coefficient squares of this MO (with basis function types $m$ and atomic centers $k$, $l$) to express the basis function contribution %$i$ for this MO:

$$\%i = \frac{\sum_k |c_{i,k}|^2}{\sum_m \sum_l |c_{m,l}|^2} \times 100\% \quad \text{with } i \in \{m\} \quad (2)$$

### 2.2 Mg(CO)$_8$

Although not a TM, our computations predict the magnesium octacarbonyl complex to be stable (see Table 1), but with about -29 kJ/mol bound significantly weaker than Ca(CO)$_8$ and [Ti(CO)$_8$]$^{2+}$. This formation energy does not account for any vibrational effects. We find that the inclusion of the zero-point energy (ZPE) correction leads to a destabilization of the complex by +31 kJ/mol ($E_f^{ZPE}$ = +2 kJ/mol), so that the complex would be slightly unstable and not observable under vibrations. However, here we are interested only in an isoelectronic non-TM reference case to study the *relative* effect of *d* basis functions on the metal carbonyl complex properties. Therefore we deem a local minimum of the electronic potential energy surface as sufficient to study general trends in vibrational frequencies, electronic structure and *relative* formation energy changes.

The full *cc*-pVQZ basis set for Mg contains six sets of *s*, five of *p*, three of *d*, two of *f* and one set of *g* functions with different radial extent. Fig. 2 illustrates the changes in average CO stretch frequencies and binding energies for different degrees of truncation of the *cc*-pVQZ basis.

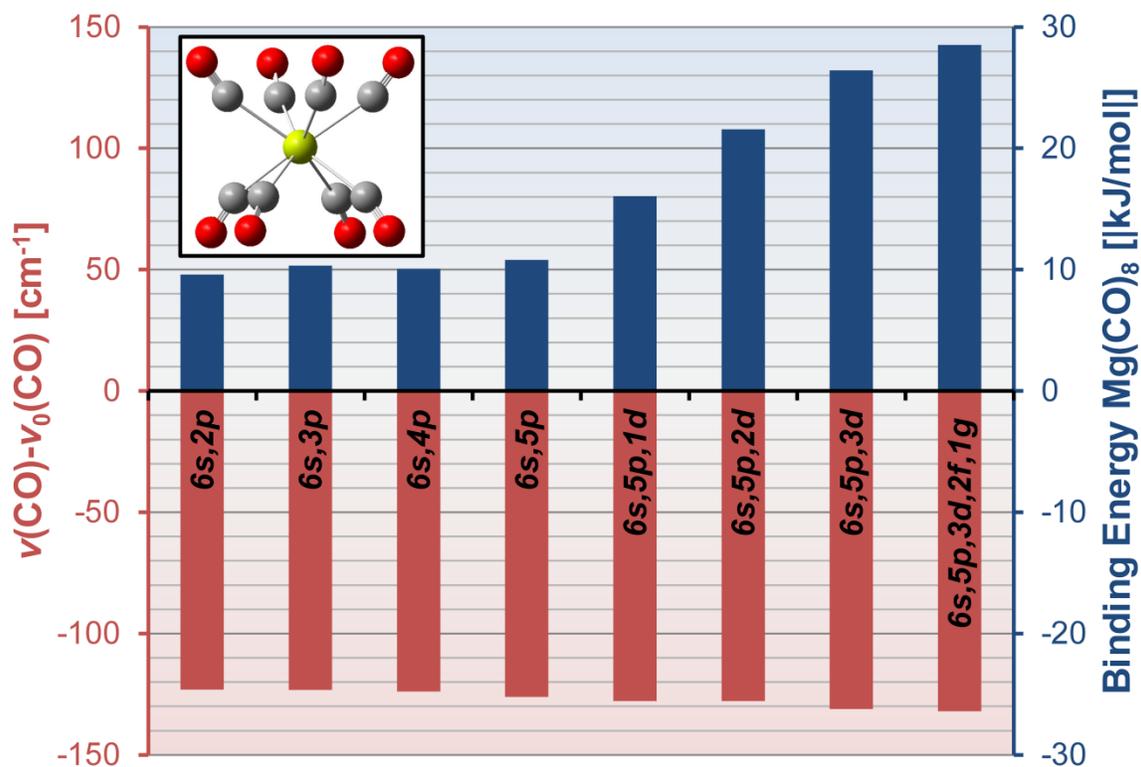

**Fig. 2** Binding energies (right axis, blue bars) and CO stretch frequencies (left axis, red bars) of the Mg(CO)$_8$ complex with changing degrees of basis set truncation. CO stretch frequencies are given as difference value to the computed free CO stretch frequency (2279 cm$^{-1}$). The complex geometry is shown in the inset.

Unsurprisingly, the binding energy increases with the basis set size, especially an increasing number of $d$ and higher-$l$ functions has a stabilizing effect on the Mg(CO)$_8$ complex. This can be mainly attributed to the beneficial effect of more diffuse functions on the weakly bound complex, as binding energies steadily increase by a significant amount with each set of $d$ functions introduced (in order of their radial extent). Besides this, as can be seen from the insets in Fig. 1, the valence orbitals change qualitatively very little by introduction of $d$ functions. For the full basis, the main contributions to the frontier orbitals are $s$-type functions with 78.8 %$s$(Mg) in the highest $a_1$ orbital and 10.17%$s$(Mg) in the lower one, in contrast to only 0.8 and 0.2%$d$(Mg), respectively, while $p$ components play no role. This in accordance to the expected low admixture of $d$ functions from high-energy Mg atomic orbitals. The remaining molecular orbitals have low contributions from the Mg central atom. In Reference 7, an energy decomposition analysis–natural orbitals for chemical valence (EDA-NOCV[34]) method was used to distinguish the type of contribution to binding of these orbitals in $D_{4d}$ octacarbonyl complexes of higher group II metals (Ca, Ba, Sr). The $a_1$ HOMO was attributed the M($d$)-to-CO π backbonding, while the remaining orbitals corresponded to either σ donation or CO polarization ($e_2$). In the case of the Mg complex investigated here, interactions between the Mg center and π* CO orbitals are enabled by overlap of one of each CO π* orbital lobes with the Mg-$s$ orbital and of π* orbital lobes among each other. It is evident from Fig. 2 and Table 1 that the CO stretch frequencies change only minimally with increasing basis size, the same is true for the ionicity of the Mg-CO interaction with Bader charges and DIs varying only slightly. *As can be seen in Table 1, there is a clear red shift (~-130 cm$^{-1}$) of the CO stretch frequencies in the Mg(CO)$_8$ complex and an increase in the C-O bond lengths, although there are no metal d functions "available" to form a π-type M-C bond.* The M-CO bond has a high degree of ionicity with a Mg Bader charge close to +1.7 (as opposed to a formal oxidation state of ±0, since the Mg-assigned highest $a_1$ is doubly occupied) and only 0.04 electrons delocalized between Mg and C basins. Destabilization of the carbon-oxygen bond can be explained by the charge transfer from the Mg central atom to the CO ligand (cf. the CO stretch frequency in a free CO$^-$ anion is lower than for the neutral species, with our computational setup a red shift of 1744 cm$^{-1}$ and a C-O distance of 1.2101 Å was obtained upon negatively charging CO). In terms of orbital interactions, this could be rationalized by the population of CO π* orbitals through a Mg-*s* σ-"backbonding" from the *formal* Mg$^0$ central atom, weakening the carbon-oxygen bond.

**2.3 Ca(CO)$_8$**

As can be seen in Table 1, the octacarbonyl calcium complex is predicted to be more than ten times more stable than the corresponding magnesium complex, which is in agreement with the observation of higher group II octacarbonyl complexes, but not of the lighter alkaline earth elements. The *cc*-pVQZ basis set for Ca contains seven sets of *s*, six of *p*, four of *d*, two of *f* and one of *g* orbitals. The change of binding energies and average CO stretch frequencies with different degrees of basis truncation is illustrated in Fig. 3.

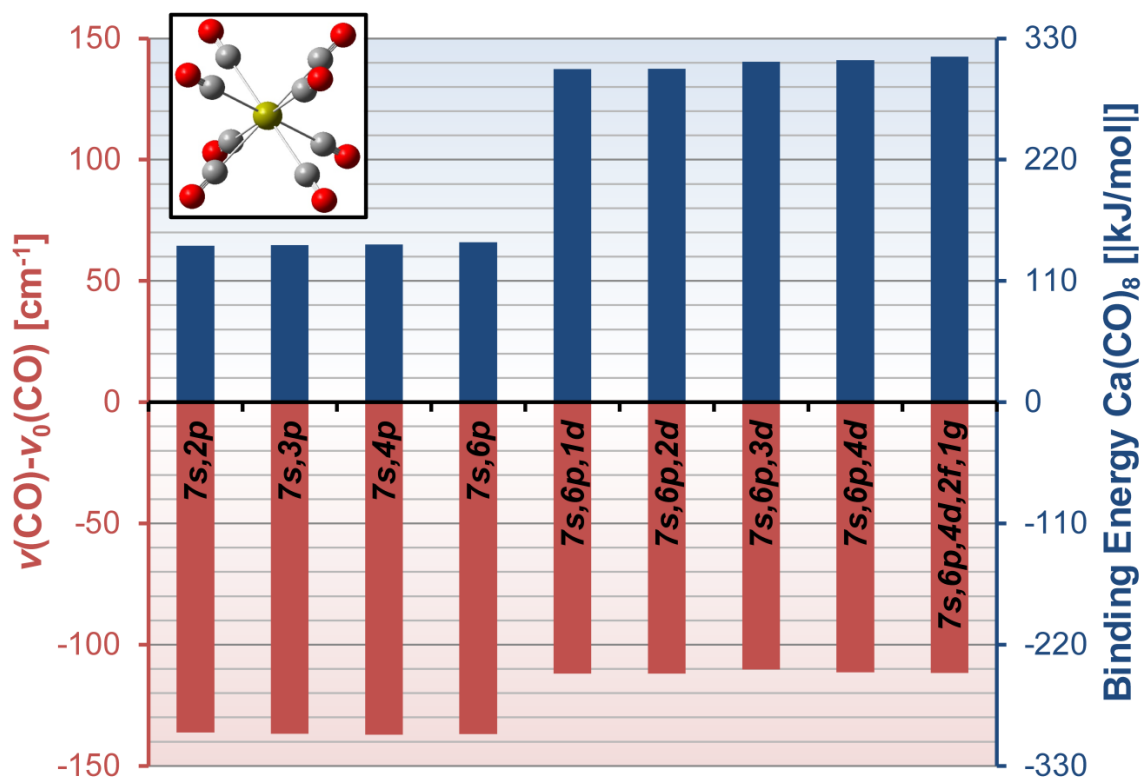

**Fig. 3** Binding energies (right axis, blue bars) and CO stretch frequencies (left axis, red bars) of the Ca(CO)$_8$ complex with changing degrees of basis set truncation. CO stretch frequencies are given as difference value to the computed free CO stretch frequency (2279 cm$^{-1}$). The complex geometry is shown in the inset.

The inclusion of *d* functions has a distinct stabilizing effect on the Ca(CO)$_8$ complex, more than doubling the complex formation energy. This occurs directly after introduction of one pair of *d* functions, while further *d* and higher polarization functions contribute only slightly to an additional stabilization of the complex (the relative change between the complex formation energies of Ca{7*s*,6*p*,1*d*}(CO)$_8$ and Ca{7*s*,6*p*,4*d*,2*f*,1*g*}(CO)$_8$ amounts to just 1%). Firstly, this shows the high importance of (specifically) metal *d* functions and the formation of M-C π bonds on carbonyl complex stability. With a truncated 7*s*, 6*p* basis set on Ca, the main contributions to Ca-CO bonding stem from the lowest-lying $a_{1g}$ and $t_{1u}$ orbitals which have 24 %*s*(Ca) and 9 %*p*(Ca), respectively, while the remaining orbitals lack any contribution from Ca-centered functions. This corresponds to a bonding purely based on "σ donation" into metal *s* and *p* orbitals, although the term might me misleading in this context, since the Bader charge analysis suggests a strong ionization of Ca and charge density transfer to the CO ligands from the central atom, instead of the opposite direction this nomenclature implies. In fact, the formal oxidation state of Ca in the complexes without *d* functions corresponds to +2, making the interactions formally a σ donation of (CO)$_8^{2-}$ without any direct π backbonding to the ligand complex. With the full *cc*-pVQZ basis set on the other hand, significant contributions of Ca *d* functions are obtained for the $t_{2g}$ and $e_g$ orbitals with 14 and 33 %*d*(Ca), respectively. As discussed previously, this is not surprising since the Ca *d* functions (from low-lying unoccupied Ca atomic orbitals) readily hybridize with the Ca *s* orbitals resulting in a lowered energy of the complex, other than in e.g. Mg(CO)$_8$ with its high-energy *d* states of the metal atom. The $e_g$ SOMOs show a clear π character (see Fig. 2), and have been previously assigned the π backbonding from metal to CO by EDA-NOCV in the work of Wu *et al.*[7] Interestingly, as can be seen in Fig. 3 and Table 1, *the red shift of the CO stretch frequencies and C-O bond elongation does not depend on the occurrence of metal-CO π bonds*. Moreover, *the CO stretch frequencies increase slightly with the formation of M-CO π bonds*. This correlates with the decreasing ionicity of the bond (see Table 1), since the *d* contributions to the SOMO and SOMO-2 level states increase the population on Ca, shifting electron density back from the CO ligands to Ca (a total of 0.3 |*e*| as found by Bader analysis), and increasing the covalency of the M-C bond as indicated by an increasing DI(M-C). In other words, the population of CO π* states decreases the CO stretch frequency, but it does not require M-CO π bonds. The latter mitigate the red shift by an effect similar to the previously reported charge self-regulation of TMs,[35] retaining some population of Ca states. We want to emphasize at this point that from the molecular orbitals of the non-*d* case no "synergistic" σ-donation/π-backbonding mechanism can be deduced, but rather an occupation of CO ligand group orbitals stemming from the combination of CO π* orbitals, causing a more severe predicted red shift than for the *d*-included case in which electrons are transferred back to $Ca^{2+}$ (+2 refers here to the formal oxidation state of Ca in the non-*d* complex, as opposed to ±0 if the full *cc*-pVQZ basis set is used on Ca). The contributions of Ca *d* states which lead to a FOS of ±0 do not however reverse a significant charge transfer from Ca to the ligands (Bader charge of +1.45 with vs +1.75 without *d* functions) – the Ca atom *is* oxidized in spite of being in a zero formal oxidation state. This highlights the fact that the FOS is not actually defined with respect to oxidation but is rather a way to define and count bonds based on non-observable functions. Failures of FOS to describe redox phenomena have been reported before;[35-40] here we see a rather severe failure.

## 2.4 $[Ti(CO)_8]^{2+}$

As a true TM carbonyl complex, the 18-electron $[Ti(CO)_8]^{2+}$ exhibits the highest stability among the three investigated complexes, more than three times more stable than the Ca and 30 times more stable than the Mg octacarbonyl complexes. The *cc*-pVQZ basis set on Ti has eight sets of *s*, seven of *p*, five of *d*, three of *f*, two of *g* and one of *h* functions. $[Ti(CO)_8]^{2+}$ is found to be a non-classical carbonyl complex, with the CO stretch frequencies being blue shifted by 86 cm$^{-1}$. As can be seen in Fig. 4, the blue shift is dependent on the presence of *d* functions and for the truncated basis sets without *d* and higher-*l* contributions a CO red shift of about 20 cm$^{-1}$ was found instead.

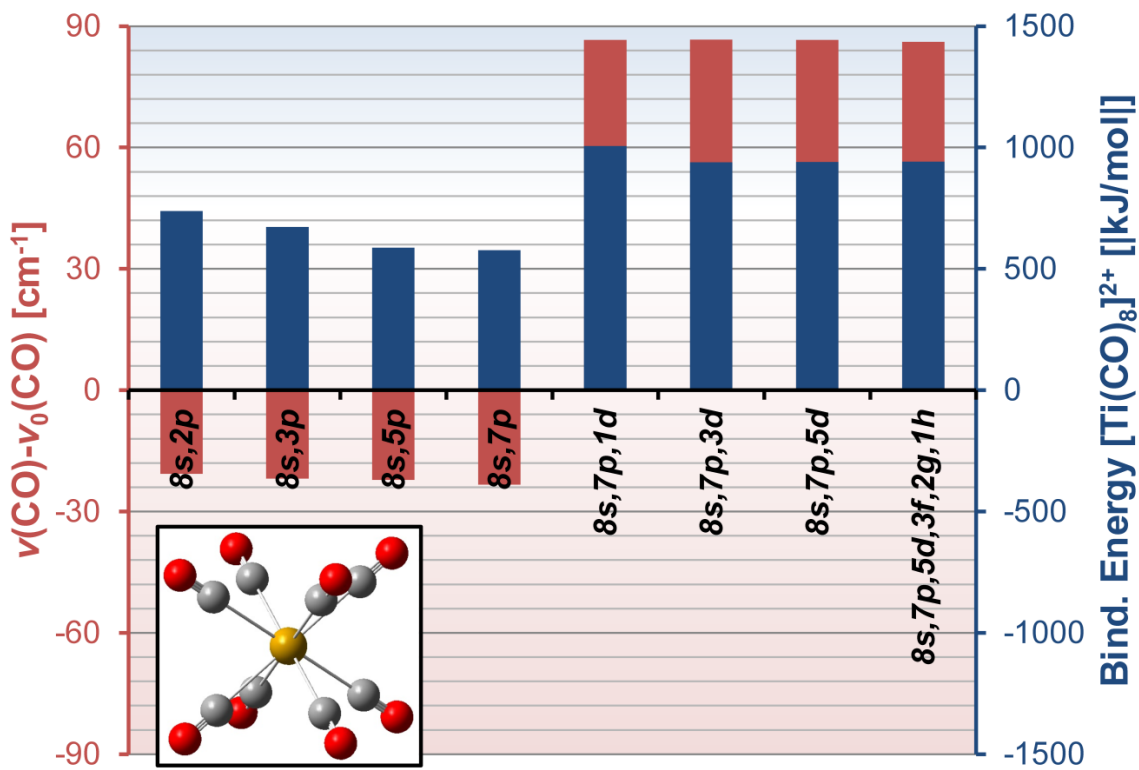

**Fig. 4** Binding energies (right axis, blue bars) and CO stretch frequencies (left axis, red bars) of the $[Ti(CO)_8]^{2+}$ complex with changing degrees of basis set truncation. CO stretch frequencies are given as difference value to the computed free CO stretch frequency (2279 cm$^{-1}$). The complex geometry is shown in the inset.

Similarly to the octacarbonyl calcium complex, the alleviation (or in this case reversal) of the CO stretch frequency red shift can be attributed to the formation of stronger metal-CO ($\pi$) bonds that lead to a back-transfer of charge density from CO to the TM, reducing the ionization of CO (and hence lowering the CO $\pi^*$ population), as can be seen from the Bader charges and DIs in Table 1. The Bader charges change from +3.5 |e| in the non-*d* case to +1.7 |e| with the full cc-pVQZ basis set (roughly corresponding to the formal oxidation states of Ti of +4 and +2 in these systems), while the M-C DI increases (indicating more electrons from both basins engaged in bonding) and the C-O DI decreases (indicating a decreasing degree of covalent bonding). When removing all *d* and higher polarization functions on Ti, the only states in Fig. 1 with contributions from the central atom are $t_{1u}$ σ-"donating" orbitals with 14 %*p*(Ti). If Ti *d* orbitals and higher-*l* functions are included, the SOMO level becomes predominantly Ti-*d* (95 %*d*(Ti)), which explains the charge transfer of almost two electrons from CO to Ti when *d* functions are included, as found from the Bader analysis. Similar to the Ca case, the formation of metal-CO π bonds leads to a substantial stabilization of the complex by a factor of roughly two. It might seem counterintuitive that the complex formation energies decrease with the inclusion of more diffuse functions of same angular momentum as can be seen in Fig. 4 for the *p*- (and to a lesser extent for the *d*-) truncated basis sets, since larger and more diffuse basis functions are normally expected to give rise to more stable binding, but this can be explained by the massive destabilization of the $Ti^{2+}$ cation reference by the (unphysical) omission of *d* (and polarization) functions which counteracts the complex stabilization (cf. eq. (1)). As previously seen for Ca, the inclusion of higher-*l* functions has only a minor effect on complex formation energies and *it is the occurrence of d-type functions that leads to a significant increase in stability, higher CO stretch frequencies and shorter C-O bonds in TM and TM-like carbonyl complexes* (see Table 1). While M-C π bonds have been already discussed as main factor of TM carbonyl complex stabilities in literature,[41, 42] the red shift of CO ligands is still commonly associated with the formation of M-C π bonds enabled by TM *d* functions. As can be seen here, this is not precisely the case, but it is rather caused by a metal-CO charge transfer which does not require M-C π bonds to take place.

## 3. Conclusions

We theoretically investigated $Mg(CO)_8$, $Ca(CO)_8$ and $[Ti(CO)_8]^{2+}$ complexes by the means of density functional theory. We used a simple basis set truncation to unambiguously discern the contributions of metal basis functions with different nodal structure (as well different radial extent) to electronic structure, energetics and geometry of these complexes and therefore the dependence of these factors on central atom orbitals. The conclusion we can draw from our investigation is threefold:

1. Central atom *d* functions play a crucial role in the stabilization of TM and TM-like carbonyl complexes relative to non-TM carbonyls. This might be the main reason for the commonly observed instability of C-bound main group metal carbonyl complexes (with exception of e.g. higher group-II elements).[43] Also, our computations suggest that $Mg(CO)_8$ would be indeed unstable under inclusion of zero-point vibrations.

2. Charge transfer from the central atom on the ligands in metal carbonyl complexes is not dependent on metal $d$ functions and the formation of M-C $\pi$ bonds which are often used to rationalize the bonding in this kind of complexes. *Charge transfer* relates here to charge density transfer, but is also applicable to formal oxidation states. Absence (e.g. in Mg) or omission (by basis set truncation) of metal $d$-function contributions leads to a metal-CO interaction with pronounced ionic character, and M-C $\pi$ bonds on the other hand lead to an increasing covalent character as can be deduced from Bader charges and delocalization indices.
3. The CO stretch frequency decrease in metal carbonyl bonds is not dependent on $d$ functions. Presence of $d$ functions and the formation of M-C $\pi$ bonds increase the covalency of the metal-CO interaction and lead to a charge transfer from CO to the metal center that reduces the (excess) charge density on CO and generally leads to stronger C-O bonds than in the more ionic case without metal $d$ functions. We cannot deduce a $\sigma$-bonding/$\pi$-backbonding mechanism in TM carbonyl complexes in the framework chosen by us.

We want to emphasize at this point that our approach is fundamentally different from previous investigations on metal-CO bonding, since we do not try to *simultaneously* differentiate between the contributions of different orbital types within the same system, but rather contrast the properties of an electronically relaxed system with significant $d$ function influence with an (artificial) electronically relaxed reference system that lacks this contributions. We argue that this approach might be physically more meaningful since it does not require the analysis of orbitals themselves but rather compares well-defined properties (with the tradeoff of not so well-defined systems). We also want to point out that, in accordance to previous findings,[14] we are able to reproduce the correlation between CO stretch frequencies and real-space charge transfer (here in the form of Bader charges and DIs), which once more elegantly provides an avenue to rationalize even minor modulations of CO frequencies that the concept of formal oxidation states cannot provide[36] and seems in general better applicable for the explanation of phenomena.[37-39] The formal oxidation state of the metal in $Mg(CO)_8$ stays $\pm 0$, switches from +2 to $\pm 0$ in $Ca(CO)_8$ and switches from +4 to +2 in $[Ti(CO)_8]^{2+}$ without and with inclusion of metal $d$ functions, respectively, implying that the character of the CO ligand changes from charge acceptor (without metal $d$ functions and $\pi$ backbonding) to neutral $\sigma$ donor (with metal $d$ functions and backbonding). We want to once more highlight the quantitative and even qualitative discrepancy of formal oxidation states and real-space charge transfer, with the latter being the more sensitive (and physically well-defined) measure to match experimental observations related to charge transfer in a meaningful manner. This is exemplified well in the case of Ca(CO)$_8$, where a zero-FOS metal shows a severe depletion of charge density around the nucleus and, moreover, a basis set truncation that decreases the formally assigned number of metal electrons by the number of two affects the charge density distribution only slightly ($\Delta q$ = 0.35 |$e$|), highlighting not only the quantitative but severe qualitative deficiency of the FOS approach to rationalization of redox phenomena.

## Acknowledgements

This work was supported by the Ministry of Education of Singapore (Grant No. MOE2015-T2-1-011).

# *Supporting Information*

**Basis Set Test**

In Figs. S1 and S2 the total energies and binding energies of the Mg(CO)$_8$ and Ca(CO)$_8$ complexes are shown with increasing size of the *cc*-pV*X*Z basis set (with *X*=D,T,Q,5), respectively.[1-4] As can be seen, energies are mostly converged form the triple-zeta level on. As a compromise between the additional accuracy the *cc*-pV5Z basis provides and computational feasibility, the *cc*-pVQZ basis set was chosen. Deviations between quadruple-zeta and quintuple-zeta level are less than 0.002 $E_h$ (5.3 kJ/mol) towards stronger binding in both cases, which does not affect the conclusions drawn in the main text. The inclusion of more diffuse functions in form of an augmented *cc*-pVQZ basis set brought an additional gain in accuracy in the Mg(CO)$_8$ case, but since the aug-*cc*-pVQZ basis set is not available for Ca, and the results with *cc*-pVQZ on Ca and aug-*cc*-pVQZ on C and O are deviating more from the *cc*-pV5Z benchmark than with the *cc*-pVQZ basis on all atoms, no augmented basis sets were used.

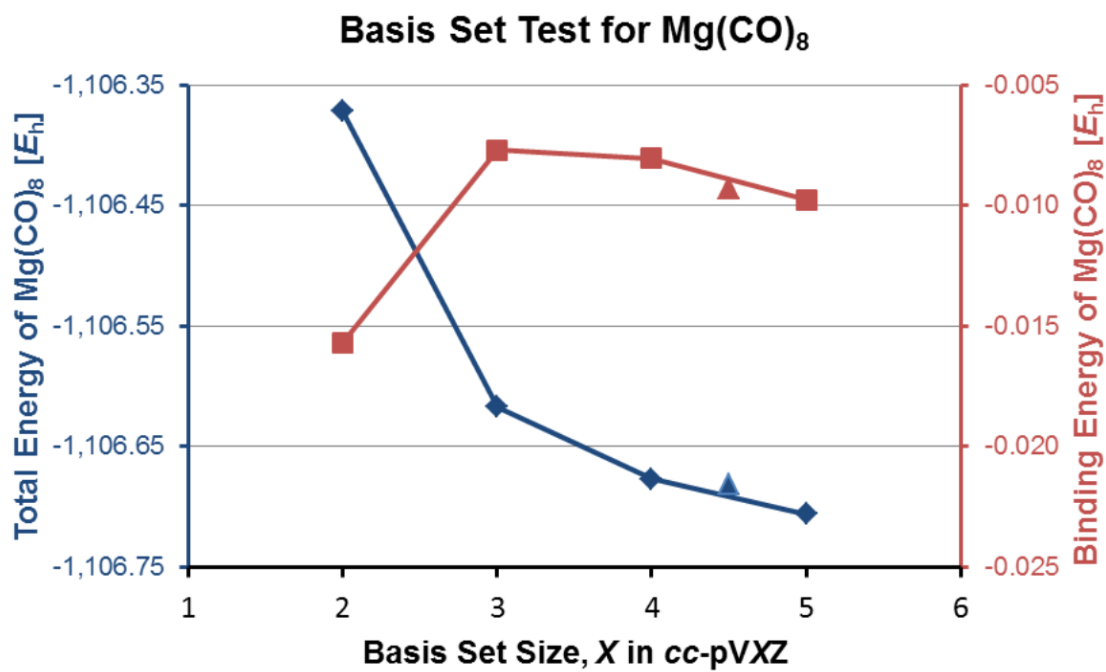

**Fig. S1** Changes in total (blue, diamonds) and binding energy (red, squares) of the Mg(CO)$_8$ complex with increasing *cc*-pV*X*Z basis set size. *X* denotes the valence-zeta and varies between D (i.e. 2) and 5. The data point at *X*=4.5 (triangle) denotes values obtained with an aug-*cc*-pVQZ

basis set for Mg(CO)$_8$ or *cc*-pVQZ on Ca/aug-*cc*-pVQZ on C and O for Ca(CO)$_8$ (since no corresponding augmented basis set for Ca is available).

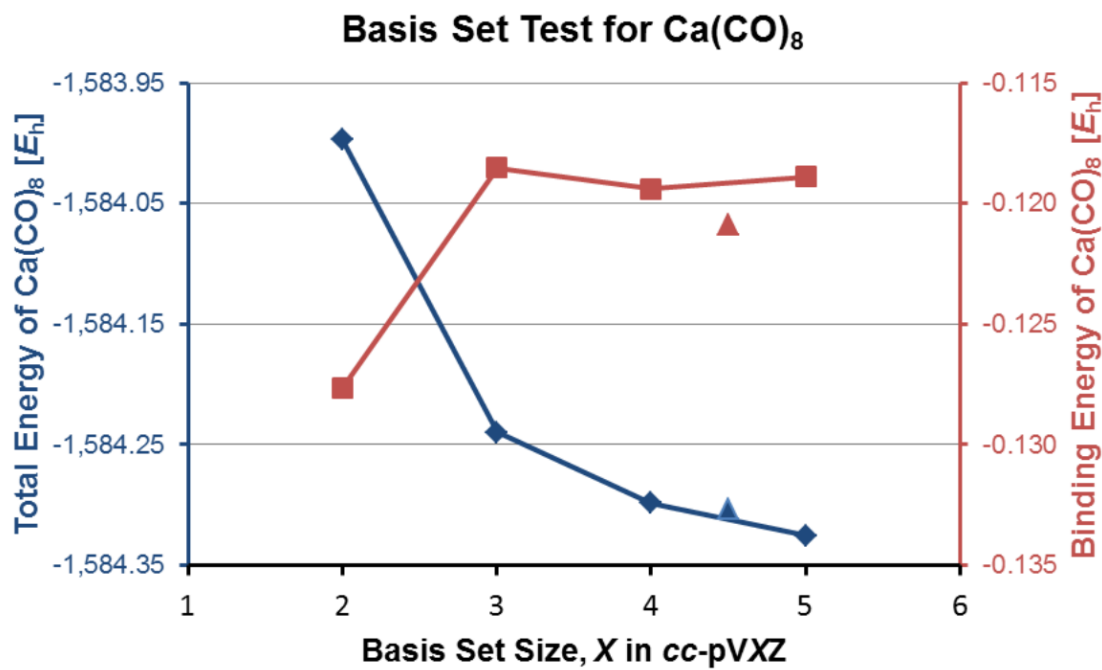

**Fig. S2** Changes in total (blue, diamonds) and binding energy (red, squares) of the Ca(CO)$_8$ complex with increasing *cc*-pV*X*Z basis set size. *X* denotes the valence-zeta and varies between D (i.e. 2) and 5. The data point at *X*=4.5 (triangle) denotes values obtained with an *cc*-pVQZ on Ca/aug-*cc*-pVQZ on C and O for Ca(CO)$_8$ (since no corresponding augmented basis set for Ca is available).

In Fig. S3, the average CO stretch frequencies in Mg(CO)$_8$ and Ca(CO)$_8$ are shown with increasing size of the *cc*-pV*X*Z basis set (with *X*=D,T,Q,5). The average CO stretch frequencies converge quickly and deviate only within 1 cm$^{-1}$ between *cc*-pVQZ and *cc*-pV5Z for these complexes, which is sufficiently precise for the conclusions drawn in the main text.

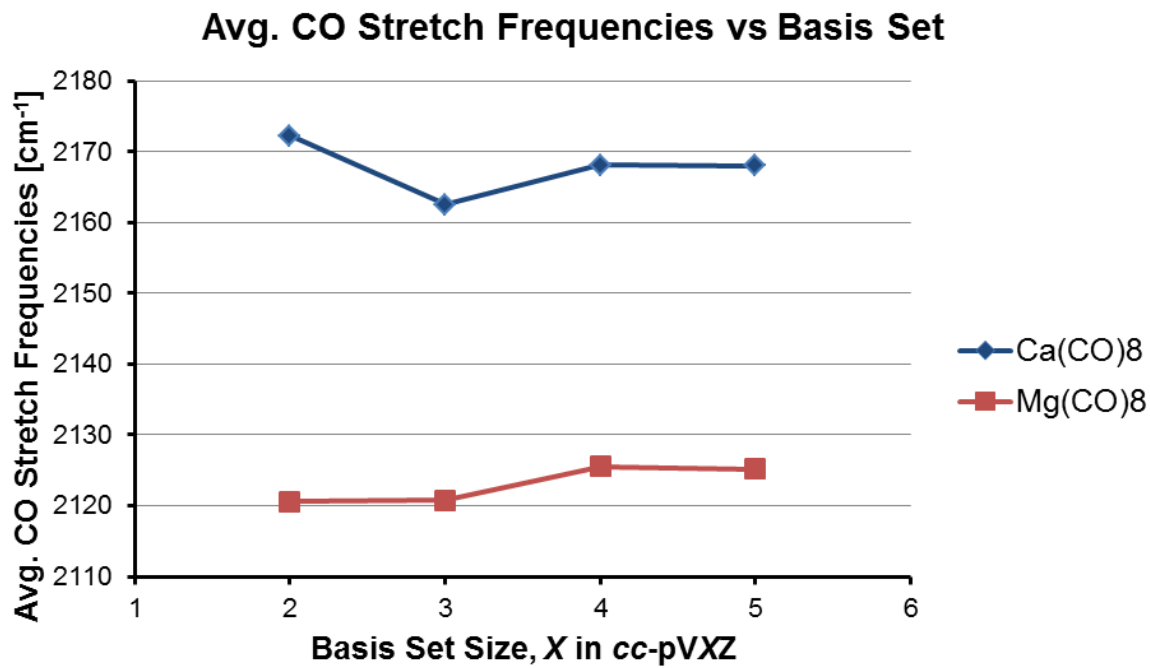

**Fig. S3** Average CO stretch frequencies in the Mg(CO)$_8$ (red, squares) and Ca(CO)$_8$ (blue, diamonds) complexes with increasing *cc*-pV*X*Z basis set size. *X* denotes the valence-zeta and varies between D (i.e. 2) and 5.

## Dispersive and Basis Set Superposition Error Effects

In Table S1 the total energy changes with an additional dispersion correction as proposed by Grimme *et al.* (DFT-D3)[5] for the three investigated complexes (with full *cc*-pVQZ basis and M06-2X functional[6] as implemented in the program package *Gaussian 16*[7]) are listed. Table S1 also includes the basis set superposition error[8,9] for the three complexes with two fragments (Ca and (CO)$_8$), indicating the magnitude of overbinding introduced by mutual basis function borrowing. Both values are small in comparison to the binding energies of each complex (see main text).

**Table S1** Dispersion correction with DFT-D3 and basis set superposition error (BSSE) for the three investigated complexes Mg(CO)$_8$, Ca(CO)$_8$ and [Ti(CO)$_8$]$^{2+}$ with M06-2X/*cc*-pVQZ.

| Complex | $E$(DFT-D3)-$E$(DFT) [kJ/mol] | BSSE(Ca/(CO)$_8$) [kJ/mol] |
|---|---|---|
| Mg(CO)$_8$ | -2.23 | 3.28 |
| Ca(CO)$_8$ | -2.47 | 7.62 |
| [Ti(CO)$_8$]$^{2+}$ | -2.44 | 5.40 |

## Influence of Complex Geometry

In Table S2 the Bader charges, CO red shifts, complex formation energies and delocalization indices for the electronically stable $^3O_h$ and $^1D_{4d}$ states of Mg(CO)$_8$ and Ca(CO)$_8$/[Ti(CO)$_8$]$^{2+}$, respectively, are listed as obtained at the M06-2X/*cc*-pVQZ level.

**Table S2** Metal Bader charges, CO stretch frequency changes (relative to free CO), Complex formation energies and delocalization indices for Mg(CO)$_8$ ($^3O_h$), Ca(CO)$_8$ ($^1D_{4d}$) and [Ti(CO)$_8$]$^{2+}$ ($^1D_{4d}$) complexes obtained with M06-2X/*cc*-pVQZ.

| | Mg(CO)$_8$ ($^3O_h$) | Ca(CO)$_8$ ($^1D_{4d}$) | [Ti(CO)$_8$]$^{2+}$ ($^1D_{4d}$) |
|---|---|---|---|
| q(M) [\|$e$\|] | +1.76 | +1.44 | +1.73 |
| $v_0$(CO)-$v$(CO)[cm$^{-1}$] | -154 | -112 | +74 |
| $E_f$ [kJ/mol] | -21.1 | -284.5 | -894.1 |
| DI(M-C) | 0.035 | 0.082 | 0.158 |
| DI(C-O) | 0.815 | 0.826 | 0.868 |

## Complex Geometries

In Tables S2 - S4, the optimized complex geometries for Mg(CO)$_8$, Ca(CO)$_8$ and [Ti(CO)$_8$]$^{2+}$ are listed in Cartesian coordinate representation as obtained with M06-2X/*cc*-pVQZ and convergence thresholds of 10$^{-6}$ E$_h$ for the total energy and 1.5·10$^{-5}$ E$_h$/a$_0$ for the interatomic forces.

**Table S3** Optimized complex geometry in Cartesian coordinates of Mg(CO)$_8$ with M06-2X/*cc*-pVQZ and convergence thresholds of $10^{-6}$ $E_h$ for the total energy and $1.5 \cdot 10^{-5}$ $E_h/a_0$ for the interatomic forces.

| Atom | Coordinates [Å] | | |
| --- | --- | --- | --- |
| | X | Y | Z |
| Mg | 0.000000 | 0.000000 | 0.000000 |
| C | 0.000000 | 1.971875 | 1.317546 |
| O | 0.000000 | 2.664534 | 2.207685 |
| C | -1.971875 | 0.000000 | 1.317546 |
| O | -2.664534 | 0.000000 | 2.207685 |
| C | 1.971875 | 0.000000 | 1.317546 |
| O | 2.664534 | 0.000000 | 2.207685 |
| C | -1.394326 | -1.394326 | -1.317546 |
| O | -1.884110 | -1.884110 | -2.207685 |
| C | 1.394326 | -1.394326 | -1.317546 |
| O | 1.884110 | -1.884110 | -2.207685 |
| C | 1.394326 | 1.394326 | -1.317546 |
| O | 1.884110 | 1.884110 | -2.207685 |
| C | 0.000000 | -1.971875 | 1.317546 |
| O | 0.000000 | -2.664534 | 2.207685 |
| C | -1.394326 | 1.394326 | -1.317546 |
| O | -1.884110 | 1.884110 | -2.207685 |

**Table S4** Optimized complex geometry in Cartesian coordinates of Ca(CO)$_8$ with M06-2X/*cc*-pVQZ and convergence thresholds of $10^{-6}$ $E_h$ for the total energy and $1.5 \cdot 10^{-5}$ $E_h/a_0$ for the interatomic forces.

| Atom | Coordinates [Å] | | |
| --- | --- | --- | --- |
| | X | Y | Z |
| Ca | 0.000000 | 0.000000 | 0.000000 |
| C | 1.501332 | 1.501332 | 1.501332 |
| O | 2.151550 | 2.151550 | 2.151550 |

| Atom | | | |
|---|---|---|---|
| C | 1.501332 | -1.501332 | 1.501332 |
| O | 2.151550 | -2.151550 | 2.151550 |
| C | 1.501332 | 1.501332 | -1.501332 |
| O | 2.151550 | 2.151550 | -2.151550 |
| C | -1.501332 | -1.501332 | -1.501332 |
| O | -2.151550 | -2.151550 | -2.151550 |
| C | -1.501332 | 1.501332 | -1.501332 |
| O | -2.151550 | 2.151550 | -2.151550 |
| C | -1.501332 | 1.501332 | 1.501332 |
| O | -2.151550 | 2.151550 | 2.151550 |
| C | 1.501332 | -1.501332 | -1.501332 |
| O | 2.151550 | -2.151550 | -2.151550 |
| C | -1.501332 | -1.501332 | 1.501332 |
| O | -2.151550 | -2.151550 | 2.151550 |

**Table S5** Optimized complex geometry in Cartesian coordinates of $[Ti(CO)_8]^{2+}$ with M06-2X/$cc$-pVQZ and convergence thresholds of $10^{-6}$ $E_h$ for the total energy and $1.5 \cdot 10^{-5}$ $E_h/a_0$ for the interatomic forces.

| Atom | Coordinates [Å] | | |
|---|---|---|---|
| | **X** | **Y** | **Z** |
| Ti | 0.000000 | 0.000000 | 0.000000 |
| C | 1.441580 | 1.441580 | 1.441580 |
| O | 2.082028 | 2.082028 | 2.082028 |
| C | 1.441580 | -1.441580 | 1.441580 |
| O | 2.082028 | -2.082028 | 2.082028 |
| C | 1.441580 | 1.441580 | -1.441580 |
| O | 2.082028 | 2.082028 | -2.082028 |
| C | -1.441580 | -1.441580 | -1.441580 |
| O | -2.082028 | -2.082028 | -2.082028 |
| C | -1.441580 | 1.441580 | -1.441580 |
| O | -2.082028 | 2.082028 | -2.082028 |
| C | -1.441580 | 1.441580 | 1.441580 |
| O | -2.082028 | 2.082028 | 2.082028 |
| C | 1.441580 | -1.441580 | -1.441580 |
| O | 2.082028 | -2.082028 | -2.082028 |
| C | -1.441580 | -1.441580 | 1.441580 |
| O | -2.082028 | -2.082028 | 2.082028 |